\def\year{2022}\relax
\documentclass[letterpaper]{article} %
\usepackage{aaai22}  %
\usepackage{times}  %
\usepackage{helvet} %
\usepackage{courier}  %
\usepackage[hyphens]{url}  %
\usepackage{graphicx} %
\usepackage{natbib}  %
\usepackage{caption} %
\usepackage{amsmath}
\usepackage[usenames,dvipsnames]{xcolor}
\usepackage[nameinlink,capitalize]{cleveref}
\usepackage{booktabs}
\usepackage{adjustbox}
\usepackage{subcaption}
\usepackage{soul}
\usepackage{multirow}
\usepackage{makecell}

\usepackage{rotating}
\usepackage{sistyle}
\SIthousandsep{,}
\definecolor{navy}{rgb}{0.1, 0.1, 0.8}
\definecolor{gray}{rgb}{0.4, 0.4, 0.4}
\definecolor{olive}{rgb}{0.1, 0.5, 0.1}
\definecolor{ruby}{rgb}{0.8, 0.1, 0.3}
\definecolor{darkpastelgreen}{rgb}{0.01, 0.75, 0.24}
\definecolor{celestialblue}{rgb}{0.29, 0.59, 0.82}
\definecolor{coral}{rgb}{1.0, 0.5, 0.31}
\definecolor{Goldenrod}{rgb}{0.8,0.8,0}

\usepackage[colorinlistoftodos,prependcaption,textsize=tiny,textwidth=15mm]{todonotes}
\usepackage{xspace}

\newcommand{\eat}[1]{}
\newcommand{\rev}[1]{{#1}}

\newcommand{\revK}[1]{{#1}}
\newcommand{\NOTE}[2]{}
\newcommand{\note}[1]{}
\newcommand{\editnote}[2][1=]{}
\newcommand{\nb}[1]{}
\newcommand{\mar}[1]{}
\newcommand{\prev}[1]{}
\newcommand{\TODO}[2]{}

\newcommand{\titlename}{Slipping to the Extreme: A Mixed Method to Explain How Extreme Opinions Infiltrate Online Discussions}
\title{\titlename}

\author {
    Quyu Kong,\textsuperscript{\rm 1,2}
    Emily Booth,\textsuperscript{\rm 1}
    Francesco Bailo,\textsuperscript{\rm 1}
    Amelia Johns,\textsuperscript{\rm 1}
    Marian-Andrei Rizoiu\textsuperscript{\rm 1,2} \\
}
\affiliations {
    \textsuperscript{\rm 1} University of Technology Sydney, \\
    \textsuperscript{\rm 2} Australian National University \\
    quyu.kong@uts.edu.au, emily.booth@uts.edu.au, francesco.bailo@uts.edu.au, amelia.johns@uts.edu.au, marian-andrei.rizoiu@uts.edu.au
}

\begin{document}

\maketitle

\begin{abstract}
Qualitative research provides methodological guidelines for observing and studying communities and cultures on online social media platforms. However, such methods demand considerable manual effort from researchers and can be overly focused and narrowed to certain online groups. This work proposes a complete solution to accelerate the qualitative analysis of problematic online speech, focusing on opinions emerging from online communities by leveraging machine learning algorithms. First, we employ qualitative methods of deep observation for understanding problematic online speech. This initial qualitative study constructs an ontology of problematic speech, which contains social media postings annotated with their underlying opinions. The qualitative study dynamically constructs the set of opinions, simultaneous with labeling the postings. Next, we use keywords to collect a large dataset from three online social media platforms (Facebook, Twitter, and Youtube). Finally, we introduce an iterative data exploration procedure to augment the dataset. It alternates between a data sampler --- which balances exploration and exploitation of unlabeled data --- the automatic labeling of the sampled data, the manual inspection by the qualitative mapping team, and, finally, the retraining of the automatic opinion classifiers. We present both qualitative and quantitative results.
First, we show that our human-in-the-loop method successfully augments the initial qualitatively labeled and narrowly focused dataset and constructs a more encompassing dataset.
Next, we present detailed case studies of the dynamics of problematic speech in a far-right Facebook group, exemplifying its mutation from conservative to extreme. 
Finally, we examine the dynamics of opinion emergence and co-occurrence, and we hint at some pathways through which extreme opinions creep into the mainstream online discourse.
\end{abstract}

\section{Introduction}

In 2020, the COVID-19 pandemic alerted the world to complex issues that arise from social media platforms circulating user-generated misinformation, hate speech, and conspiracy theories~\citep{posetti2020disinfodemic}.
Such forms of problematic information \citep{jack2017lexicon} have been studied before, with the influence of disinformation campaigns on elections~\citep{Kim2019}, disaster management~\citep{rajdev2015fake} and other global public health promotions~\citep{bode2018see} being recorded in the literature. 
To date, there exist three primary types of methods for addressing problematic information.
The first type \revK{concentrates} on large-scale monitoring of social media datasets to detect inauthentic accounts (bots and trolls)~\citep{kong2020modeling,kong2020describing,Ram2021a}, coordinated disinformation campaigns~\citep{Rizoiu2018a} and detect the usage of hate speech in social media~\citep{Rizoiu2019}.
The second group aims to understand which platforms, users, and networks contribute to the ``infodemic''~\citep{smith2019mapping,bruns2020covid19,colley2020challenges}.
The third group uses computational modeling to predict future pathways and how the information will spread~\citep{molina2019fake}. 
These studies provide valuable insights into understanding how problematic information spreads and detecting which sources are reshared frequently and by which accounts.
Though the first and third research approaches offer a breadth of knowledge and understanding, there are limitations --- they often have less to say about why certain opinions and views gain traction with vulnerable groups and online communities. Qualitative research methods are well placed to address this gap.

Qualitative methods provide rich, contextual insights into the social beliefs, values, and practices of online communities, which shape how information is shared and how opinions are formed~\citep{boyd2010social,baym2015personal,johns2020will,wu2021cross}.
This is also fundamental to understanding how and why certain opinions and information sources scale to encompass large segments of the online society~\citep{bailo2020online,bruns2020covid19}.
However, a common criticism of qualitative research is that the in-depth knowledge comes at the expense of generating insights of limited representativeness and weak robustness of the findings.
Therefore, there is a gap between the depth of insight gained from ethnographic and qualitative approaches and the breadth of knowledge gained from computational methods from data science.

This paper aims to fill this gap by proposing a mixed-method approach that brings together qualitative insights, large-scale data collection, and human-in-the-loop machine learning approaches.
We apply our method to map both in-depth and in-breadth the problematic information around four topics: \textit{2019-20 Australian bushfire season}, \textit{Climate change}, \textit{COVID-19}, and \textit{Vaccination} on three social media platforms (Facebook, Twitter and Youtube).
Specifically, this work addresses three open questions concerning applying machine learning and qualitative research in analyzing problematic online speech.

\begin{figure}[!tbp]
	\centering
    \begin{subfigure}{0.47\textwidth}
		\includegraphics[width=\textwidth]{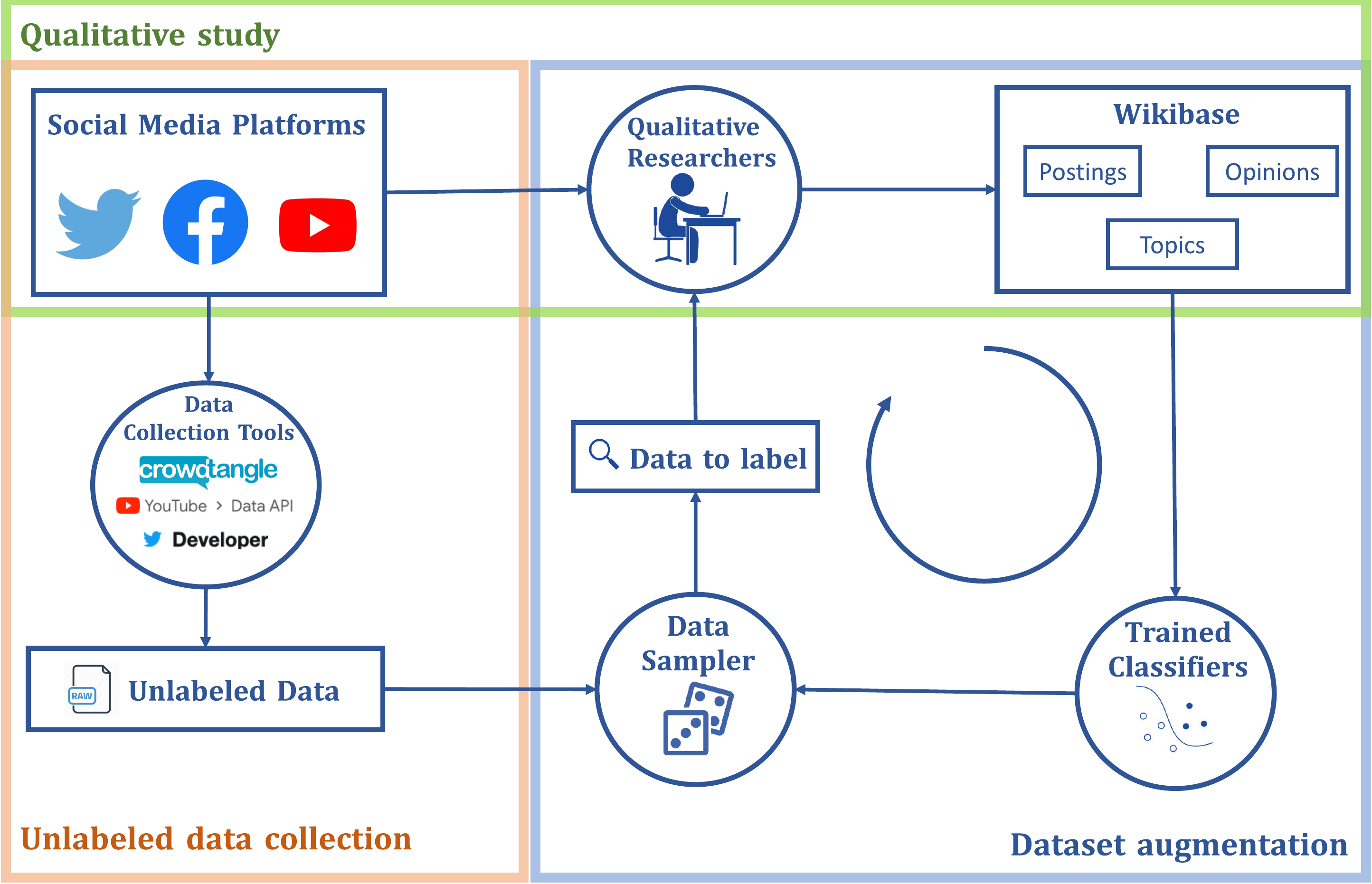}
	\end{subfigure}
	\caption{
        The pipeline of machine learning accelerated qualitative research where the human-in-the-loop machine learning algorithms are employed for dataset augmentation.
	}
	\label{fig:teaser}
\end{figure}

The first research question emerges naturally from the gap: \textbf{can we leverage both qualitative and quantitative analysis for studying problematic online speech?}
To address the challenge, we present a complete solution that bridges and facilitates both analyses (shown in \Cref{fig:teaser}). We first build a platform based on an open-source tool, Wikibase, where qualitative and quantitative analysis is conducted. 
It enables constructing an ontology of problematic online speech by performing the qualitative study, which labels data by topics and builds the vocabulary of opinions simultaneously. 
We then collect large-scale raw data using the uncovered vocabulary. 
Next, we employ machine learning algorithms to augment the data labeling process in a human-in-the-loop setting. 
Finally, we show a sample thematic and discourse analysis from the qualitative study focused on two examples of Facebook posts and comments from a far-right public Facebook group, and the quantitative outcome with measurements and statistics of the produced vocabulary.

The second question concerns the scaling of the qualitative approaches.
Such approaches require the team to observe, record and collect online discussions. 
One needs to manually identify online communities where problematic speech occurs and annotate pieces of texts with their underlying opinions.
Therefore, this in-depth exploration faces two challenges --- a significant amount of effort from researchers and the introduction of human bias in the process of collecting information \citep{dixon2016computer}.
While machine learning is known to help data exploration at scale \citep{lin2012large}, a question remains:
\textbf{can we accelerate qualitative research and observations of online behavior with machine learning algorithms?}
We tackle this challenge by adopting the state-of-the-art text classification algorithm, RoBERTa \citep{vaswani2017attention,liu2019roberta}, with a human-in-the-loop learning setting.
We first train the classifiers to identify problematic speech on postings annotated by the qualitative researchers. 
Next, we deploy three strategies to select unlabeled data. 
The active learning~\citep{settles2012active} strategy selects the data for which the classifiers are most uncertain.
The top-confidence strategy selects data that classifiers are most certain about.
The third strategy --- the random strategy --- randomly samples from unlabeled data. 
The qualitative researchers then label the sampled data, introduce the newly labeled data in the ontology, and repeat the procedure iteratively until the predictive performance converges.

The last research question relates to applying the qualitative mapping at scale and analyzing the dynamics of problematic opinions.
The question is \textbf{can we track the dynamics of problematic opinions from online discussions using unlabeled data?}
To answer this question, we leverage the opinion classifiers that we build on the augmented labeled set.
First, we automatically label the opinions in a large set of postings spanning \rev{more than a year, from July 2019 until October 2020.}
This allows us to apply the qualitative-defined coding schema to a significantly larger sample of postings, therefore reducing the unavoidable selection bias of the qualitative study.
\revK{
It also offers \rev{a longitudinal quantitative approach to studying how fringe opinions capture attention via co-occurrence with mainstream opinions.}
We build a network of opinion co-occurrences from the machine-labeled dataset.
We make several observations: first, we investigate the evolution of opinion co-occurrences and highlight three types of dynamics (stable, increasing, and decreasing co-occurrence weight);
next, we examine the conspiracy opinions in the network via centrality measures and identify \rev{their spikes followed by decreasing centrality due to} the efforts of media in debunking them; 
last, we observe that conspiracy opinions are frequently\rev{ rationalized and popularized by} embracing core opinions (e.g., ``Climate change isn't real'').}

The main contributions of this work include:
\begin{itemize}
    \item A mixed-method solution for bridging qualitative and quantitative analysis, including the hosting platform (Wikibase), the initial qualitative study, the unlabeled data collection and the dataset augmentation with machine learning algorithms.
    \item A dataset augmentation procedure that merges qualitative approaches with machine-learning-based human-in-the-loop data augmentation methods.
    \item Case studies of the evolution of problematic online speech in an Australian far-right Facebook group.
    \item Analysis of problematic opinions emergence and co-occurrence by applying quantitative methods on the collected raw data.
\end{itemize}

\section{Methods}

This section details our methodology, which includes three distinct phases implemented sequentially (shown schematically in \Cref{fig:teaser}):
the \emph{qualitative study} (\cref{subsec:qual-study}),
the \emph{unlabeled data collection} (\cref{subsec:unlabeled-data-collection}), and
the \emph{dataset augmentation} using machine learning (\cref{subsec:dataset-augmentation}).

\subsection{Qualitative Study}
\label{subsec:qual-study}

A set of known far-right community pages served as the \emph{data entry point} of the qualitative study, after which we let ourselves guided by users' posting and linking, and recommendation algorithms.
We employed unobtrusive observation approaches to observe internet places where problematic speech occurs, create field notes of rich, qualitative data, construct a vocabulary of opinions to describe it, and gather and label data.

\subsubsection{Choice of qualitative method.}
The team was initially committed to using \emph{digital ethnography} as the \emph{methodological entry point} for studying problematic online content.
Ethnography is a research method that allows the object of the study to ``emerge through fieldwork, as the significant identities and locations unfold''~\citep{hine2015}, rather than predefining a set of users, sites, or keywords to construct the dataset.
When using this method, the researchers are involved hands-on with the participants they study -- i.e., they are visible, participate in discussions and ask questions~\citep{baym2009making}.
However, given the nature of the field and the communities studied in this project, the intrusion or participation of the researcher in community fora may have an undue influence on online discussions.
Therefore, we opted instead for a deep \emph{qualitative study} in which we undertake unobtrusive observations of conversations in public pages, forums, groups, and sites.
However, the rest of the methodology introduced in this paper would work just as well with a proper ethnographic approach.

\subsubsection{Problematic speech.}
Problematic speech is online interactions, speech, and artifacts that are inaccurate, misleading, inappropriately attributed, or altogether fabricated \citep{jack2017lexicon}.
The concept is intentionally broad to encompass concepts like misinformation, disinformation, and hate speech.
\emph{Misinformation} is a type of communication where falsehoods are unintentionally shared by users~\citep[p.~2]{jack2017lexicon}.
\emph{Disinformation} is information that is ``deliberately false and misleading'' \citep[p.~3]{jack2017lexicon} and intended to manipulate users to a particular opinion or worldview, and
\emph{hate speech} refers to ``any form of communication in which others are attacked, denigrated, or intimidated based on religion, ethnicity, gender, national origin, or another group-based trait''~\citep{warner2012detecting,hameleers2021civilized}.
Prior literature suggests an intertwining of these forms of problematic speech as efforts to denigrate outgroups are common to online disinformation campaigns.
\citet{hameleers2021civilized} argue that ``politically motivated, partisan or ideological utterances in false information, such as hate speech and incivility, may be an indicator of disinformation''.

\subsubsection{Study design.}\label{sssec:study_design}
Our qualitative study concentrates on discourses and conversations about four topics manually selected apriori: \textit{2019-20 Australian bushfire season}, \textit{Climate change}, \textit{COVID-19}, and \textit{Vaccination}.
We focus on three major online social media platforms, Facebook, Twitter, and Youtube, selected due to their large volume of discussion around the four chosen topics.
The study unfolded in four steps.
First, from December 2019 through January 2021, one team member undertook unobtrusive observation of discussions, collected field notes and digital artifacts (screenshots, linked data, photos, memes).
Second, the qualitative researcher labeled the data with topics and opinions that she inferred from the content.
Third, the collected data was independently double-coded by a second team member, obtaining an inter-annotator agreement of $81.0\%$.
Forth and last, the two coders reviewed the coded data and resolved disagreements through discussions.

To conduct our digital fieldwork, we first
selected a set of \emph{Internet places} --- Internet place is a generic term denoting where online discussions happen, e.g., Facebook groups or Youtube video comment sections.
In this study, we concentrate solely on publicly accessible places and
identify relevant places using four approaches:
\begin{itemize}
    \item \textbf{News stories identification.} We used the search engines of news content aggregators (e.g., Factiva, Media Cloud, LexisNexis) to identify news stories containing keywords related to chosen topics in the titles.
    Next, we observed the user comments on the articles.
    Finally, we searched social media for postings that mention the news articles. The keyword terms were constantly updated in these early stages of data collection and during iterative processes of coding the data, until a consolidated list was composed (shown in \Cref{tab:keywords}).
    \item \textbf{Page monitoring.} We actively monitored particular users, pages, and Facebook groups found at the previous point.
    We show the analysis of two such groups in \Cref{sec:case_studies}.
    \item \textbf{Cross-page discussion tracking.} We followed links in postings to discussions around the same topics on different Internet places, which we added to the list for tracking.
    \item \textbf{Exploiting recommendations.} We explored social media pages and accounts recommended by the platforms' recommender systems.
    While this introduces algorithmic bias in the sampling, this has been applied in prior literature \citep{woolley2016automation,woolley2018computational} to construct prospective pathways connecting like-minded users.
\end{itemize}

\subsubsection{An ontology to map online problematic speech.}
We collect and store information about four types of entities: \emph{topics}, \emph{postings}, \emph{Internet places} and \emph{opinions}.
The topics are predetermined, while the latter three emerge from the qualitative study.
Note that the postings and Internet places are \emph{data} discovered using the methodology described above, and the opinions are the \emph{vocabulary} describing the data.
Opinions are defined as ideas expressed by a user in a posting.
We construct new opinions during the qualitative study and the data augmentation phase and alter old opinions through merging or splitting.
As a result, we obtain the opinions simultaneously as the data is collected and labeled.%

Both the data (postings and Internet places) and the vocabulary (topics and opinions) are stored in an ontology, in Resource Description Framework (RDF) format~\citep{brickley1999resource}.
Each entry is a triplet linking two entities --- e.g., a posting contains an opinion, or an opinion is linked to a topic.
If, for example, a posting contains more than one opinion, we use multiple triplets, one for each relation.
We use \textit{Wikibase}\footnote{\url{https://wikiba.se/}} as the project's collaborative application for data input and exploration.
Wikibase offers a user-friendly interface to enter new information and connect to existing data (e.g., a new posting expressing an existing opinion);
a navigational tool to explore the links connecting the data; and
an API to search and access the data based on SPARQL queries.

\subsection{Unlabeled Data Collection}
\label{subsec:unlabeled-data-collection}

\begin{figure}[!tbp]
		\includegraphics[width=0.45\textwidth,page=1]{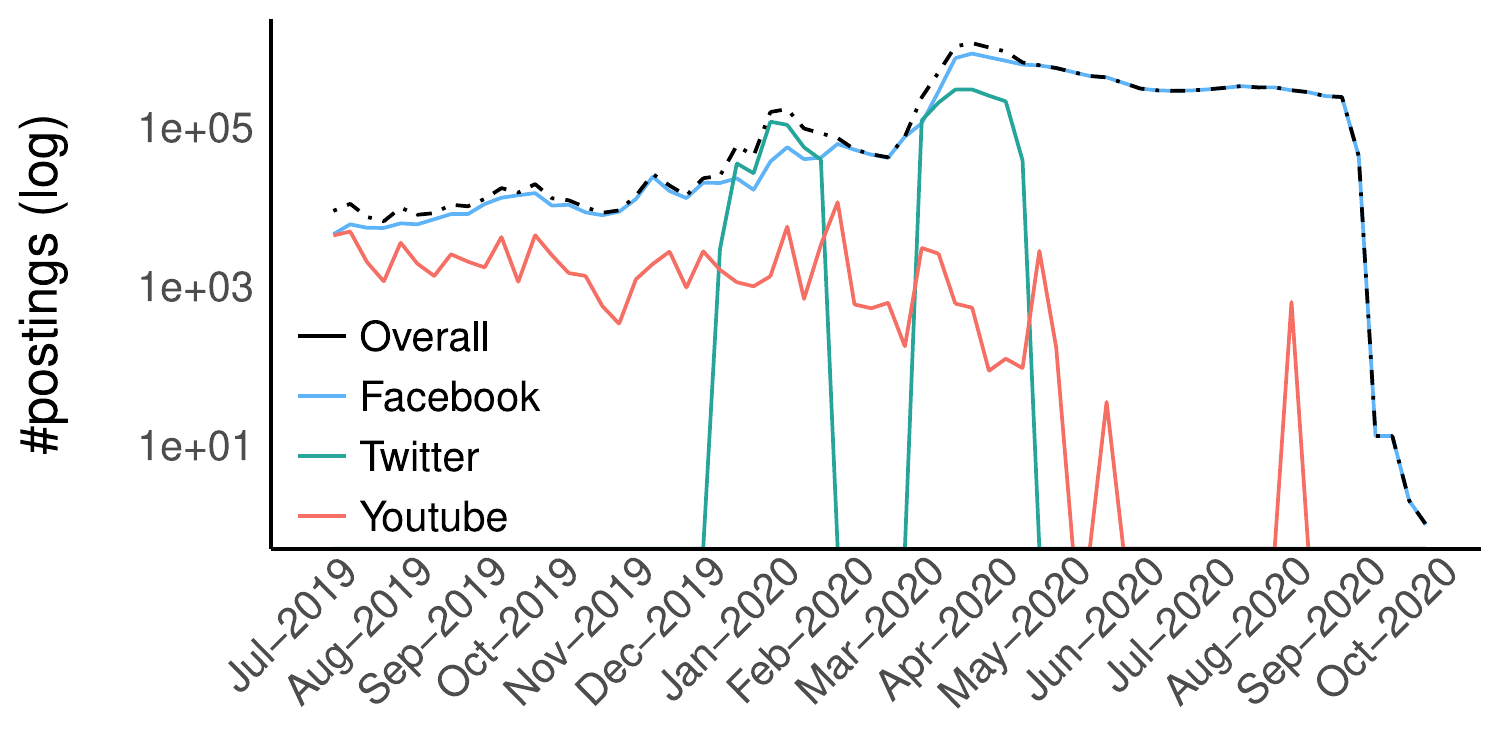}
	\caption{
		Weekly volumes of collected postings overall (dashed) and from Facebook, Twitter and Youtube (solid).
	}
	\label{fig:dataset_profiling}
\end{figure}

\begin{table}[!tbp]
    \fontsize{9pt}{10pt}\selectfont
        \begin{tabular}{@{}ll@{}}
            \toprule
            Topics & Selected keywords \\ \midrule
            \begin{tabular}[c]{@{}l@{}}2019-20 Australian\\ bushfire season,\\ Climate change\end{tabular} & \begin{tabular}[c]{@{}l@{}}bushfire,  australian fires, arson,\\ scottyfrommarketing, liarfromtheshiar,\\ australiaburns, australiaburning,\\ itsthegreensfault, backburning,\\ back burning, climate change,\\ climate mergency, climate hoax,\\ climate crisis, climate action now\end{tabular} \\ \midrule
            \begin{tabular}[c]{@{}l@{}}Covid-19,\\ Vaccination\end{tabular} & \begin{tabular}[c]{@{}l@{}}covid, coronavirus, covid-19, pandemic,\\ world health organization, vaccine,\\ social distancing, quarantine, plandemic,\\ chinavirus, wuhan, stayhome,\\ MadeinChina, ChinaLiedPeopleDied, 5G,\\ chinacentric\end{tabular} \\ \bottomrule
            \end{tabular}
    \caption{Selected keywords for topics}\label{tab:keywords}
\end{table}
One shortcoming of qualitative studies is the limited representativeness of the gathered data.
This section describes the collection of postings at scale via keyword search.
For each of the four topics, the qualitative study identified a set of keywords (shown in \Cref{tab:keywords}).
The qualitative experts created an initial candidate set of keywords using a mixture of prior knowledge and expertise, as they have been following these topics for years in previous research~\citep{JohnsFlagging2017}.
Next, they fine-tuned the set of keywords based on their frequencies observed during the qualitative study.
Due to the overlap in the messaging between \textit{Australian bushfires} and \emph{Climate change} on one side, and \emph{Covid-19} and \emph{vaccination} on the other side, we present them in two groups.
We use these keywords to search and crawl postings and comments from Facebook (using Crowdtangle\footnote{\url{https://www.crowdtangle.com/}}) and Twitter (using the Twitter commercial APIs).
We further use a customized crawler to gather comments from specific public Facebook pages and groups.
Finally, we use the YouTube API to obtain comments from the Youtube videos mentioned in the Facebook postings.
We obtained a total of $13,321,813$ postings --- $11,437,009$ Facebook postings, $1,793,927$ tweets and $90,877$ Youtube comments.
Our dataset extends from July 2019 until October 2020. \Cref{fig:dataset_profiling} shows the weekly volumes of collected postings.
Note that, for Twitter, we acquired data relating to two time periods: December 2019 -- February 2020 (during the \textit{2019-20 Australian bushfire season}) and March--April 2020 (the starting phase of \textit{Covid-19}).

\subsection{Dataset Augmentation}
\label{subsec:dataset-augmentation}

Here, we describe the process of augmenting the labeled dataset.
The augmentation process has two mandates.
First, we want to leverage the previously collected unlabeled data to create a labeled dataset containing a more encompassing set of opinions and postings compared to the data issued from the qualitative study.
Second, given the size of our unlabeled dataset, we want to maintain the manual labeling effort as limited as possible.
We enrich the dataset iteratively.
At each iteration, we use the machine classifiers to select a batch of previously unlabelled postings which are then annotated by the experts.
We denote the labeled and unlabeled datasets as $L_i$ and $U_i$, respectively, where $i$ indicates the iteration number and $i=0$ is the initial dataset labeled via qualitative analysis.

\begin{figure}[!tbp]
	\centering
    \begin{subfigure}{0.48\textwidth}
		\includegraphics[width=\textwidth]{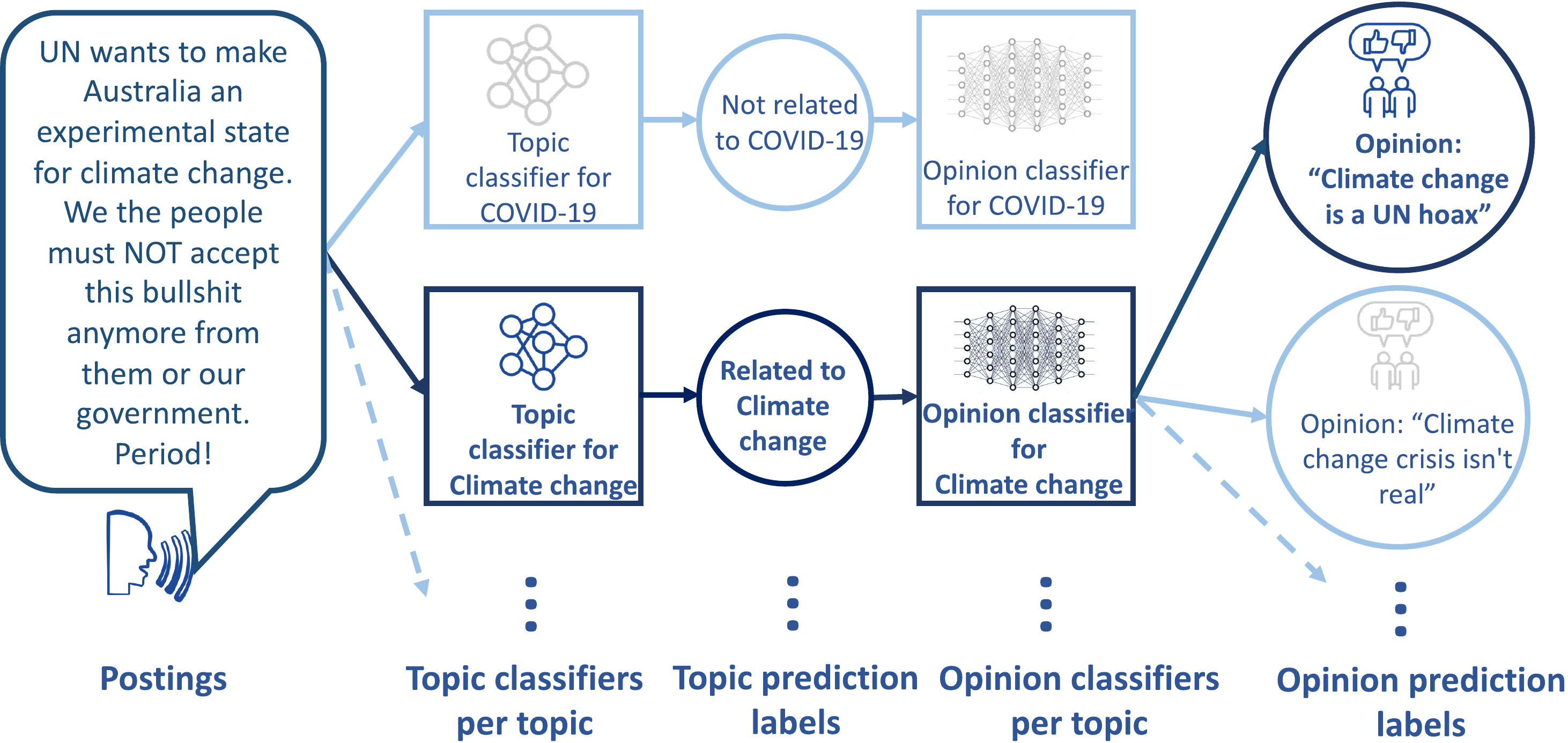}
	\end{subfigure}
	\caption{
        An example of the classification of unlabeled postings with the topic classifiers and opinion classifiers.
	}
	\label{fig:teaser-classifiers}
\end{figure}

\subsubsection{Two levels of classifiers.}
\Cref{fig:teaser-classifiers} shows our classification schema with two levels of connections:
a posting is associated with none, one or more topics and within a topic exist none, one or more opinions.
Given this hierarchy, we deploy two levels of binary classifiers.
\begin{itemize}
    \item At the first level, for a posting
          $\pmb{x}$ we construct the topic classifiers $\hat{y_t}=f_{t,i}(\pmb{x})$ ($\hat{y_t} \in \{0, 1\}$) which determine whether the posting $\pmb{x}$ is about the topic $t$, with the classifier trained on $L_i$.
          Note that we build one classifier for each topic, and a posting can be associated with multiple topics.
          It can also have no topic when $\hat{y_t} = 0, \forall t \in \{1,\dots,4\}$.
          These are off-topic postings.
    \item At the second level, we construct a multi-label opinion classifier for each topic trained with only the opinions associated with a given topic.
    Note that we train the the opinion classifiers solely after the dataset augmentation is complete as we only need the topic classifiers to perform the dataset augmentation.
\end{itemize}
We present a classification example in \Cref{fig:teaser-classifiers} where an unlabeled posting is first determined to be about \textit{Climate change} by the topic classifiers and is then tagged with the opinion ``Climate change is a UN hoax''.
We argue that the proposed scheme with two levels of classifiers is more robust to off-topic postings, as the multi-label opinion classifier is presented only with relevant postings.
Furthermore, a posting can be associated with multiple topics and opinions.

\subsubsection{Unlabeled data sampling.}\label{sssec:sampling_process}
At each iteration, we select a batch of unlabeled postings for manual annotation to augment the labeled dataset.
Within each batch, we aim to balance the \emph{exploitation} of previously labeled data (i.e., the classifiers trained at the previous iteration) and the \emph{exploration} of unlabeled data.
As unlabelled postings require first a topic label (see \Cref{fig:teaser-classifiers}), we only use the output of the topic classifiers.
We employ three strategies to select unlabelled postings at the current iteration, $X_i$
\begin{itemize}
    \item \textbf{Active learning strategy} selects for labeling the postings of which the classifiers are least certain.
    It improves classification performance by selecting unlabeled data around the decision boundary of the learned classifier~\citep{settles2012active}.
    Specifically, we adopt uncertainty sampling in our experiments where uncertainty is defined as~\citep{tran2018combining}:
      \begin{equation}
          u(\pmb{x}) = 1 - p(\hat{y} \mid \pmb{x}; f_{t,i})
      \end{equation}
    where $\hat{y}$ is the predicted label of the candidate $\pmb{x}$ under classifier $f_{t,i}$. We choose candidates with the highest uncertainty values and denote this set as $X_i^{A}$.
    \item \textbf{Top confidence strategy} chooses from unlabeled data where trained classifiers produce the highest confidence scores, i.e., $p(\hat{y} \mid \pmb{x}; f_{t,i})$. This strategy enriches our dataset with data related to the chosen topics, allowing us to deepen the qualitative study. We denote this subset as $X_i^{T}$.
    \item \textbf{Random sampling strategy} favors a completely random exploration by uniformly selecting a set of postings from the unlabeled data.
    Although there is a high likelihood of selecting off-topic postings, the strategy allows uncovering discussions of interest that may lie far from the initial qualitative analysis.
    Similar ideas have been employed in other fields --- e.g., in reinforcement learning, a probability of $\epsilon$ is usually reserved for the Q-learning algorithm to explore random actions \citep{mnih2013playing}.
    Such probability is typically small and in our experiments in \cref{sec:results}, we set the random sampling strategy to account for only $20\%$ of the sampled data.
    We denote this subset as $X_i^{R}$.
\end{itemize}

\subsubsection{Expert annotation.}
At each iteration, the same team members, who performed the qualitative analysis, label the postings returned by the sampling process (\Cref{sssec:sampling_process}).
The predicted labels from the classifiers are hidden during manual labeling.
This ensures that human decisions are not affected by algorithmic predictions.
The human experts inspect both the text and original contexts of given postings --- such as the complete discussions and other metadata (e.g., the videos from Youtube) --- before choosing an existing opinion (or constructing a new opinion) to label a posting as described in \Cref{sssec:study_design}.

\subsubsection{Iterations and convergence.}
We obtain a set of newly annotated postings at the end of each complete iteration that includes data sampling, expert annotation, and retraining the classifiers. %
For each iteration, we compute the expected generalization error via cross-validation, and we evaluate the test error on a dedicated test dataset.
The test dataset was randomly sampled from the unlabeled data and annotated before \rev{performing} the dataset augmentation.
It is kept fixed across iterations and never used in training.
We repeat the dataset augmentation process for several iterations until the convergence of two indicators:
\begin{itemize}
    \item The first indicator is the difference between cross-validation error and test set error.
    An increasingly smaller error indicates that the classifiers generalize better to the larger, unlabeled dataset.
    \item The second indicator is the gain of performance on the test set between two iterations.
    A decreasing gain between iterations shows that the marginal utility of new annotations is increasingly smaller.
\end{itemize}
The iterative process stops when an insignificant gain is made between two consecutive iterations.

\section{Dataset Augmentation Results}
\label{sec:results}
This section presents the prediction setup and results for our proposed human-in-the-loop dataset augmentation.

\begin{table}[!tbp]
        \fontsize{9pt}{10pt}\selectfont
        \begin{tabular}{rrrrr}
        \toprule
                & RF & SVM & XGBoost & RoBERTa \\
        \midrule
        Macro Accuracy & 0.791 & 0.775 & 0.779   & \textbf{0.800}   \\
        Macro F1       & 0.782 & 0.768 & 0.768   & \textbf{0.800}   \\
        \bottomrule
        \end{tabular}
    \caption{Cross-validation performance comparison of different classification models on labeled data $L_0$. Macro accuracy and F1 scores are averaged over all topics.}\label{tab:classifiers}
\end{table}

\begin{table}[!tbp]
	\label{tab:dataset}
	\centering
	\fontsize{9pt}{10pt}\selectfont
		\begin{tabular}{rrrrrrrr}
			\toprule
			&  & Aus. & Clim. & Cov. & Vac. & Off- & Total \\
			 &  & Bush- & change & 19&  & topic & unique \\
			 & & fire & & & & & \\
			 \hline
			 \multirow{3}{*}{\rotatebox{90}{\#posts}} & $L_0$ & 189 & 387 & 263 & 220 & 0 & 614 \\
			\\[-0.54em]
			 & $L_7$ & 287 & 592 & 477 & 335 & 480 & 1381 \\[0.3em] \hline
			 \\[-0.54em]
			 \multirow{3}{*}{\rotatebox{90}{\#opin.}} & $L_0$ & 16 & 31 & 29 & 22 & / & 65 \\
			\\[-0.54em]
			 & $L_7$ & 16 & 33 & 34 & 26 & / & 71 \\[0.3em] \bottomrule
			\end{tabular}
    \caption{Statistics of the labeled datasets $L_0$ and $L_7$ in topics and opinions.}
\end{table}

\subsection{Experimental Setups}

\subsubsection{Textual classifier selection.}
We predict the topics and opinions of postings using textual classifiers.
We test four such classifiers.
The first is the state-of-the-art deep learning method, RoBERTa \citep{vaswani2017attention,liu2019roberta}, which achieves the best performance.
The other three are traditional classifiers --- including Random Forest (RF) \citep{breiman2001random}, Support Vector Machine (SVM) \citep{chang2011libsvm} and XGBoost \citep{xgboost} --- which use an n-gram-based vectorial representation, where features are weighted with Term Frequency Inverse Document Frequency (TF-IDF)~\citep{rajaraman2011mining}.
We use the implementation of these algorithms from the Python libraries \textit{scikit-learn} \citep{scikit-learn} and \textit{transformers} \citep{wolf-etal-2020-transformers}.

We compare the prediction performance of these models on the $L_0$ labeled dataset (issued from the qualitative study).
We show in \Cref{tab:classifiers} the macro accuracy and F1 scores obtained via $5$-fold cross-validations.
The hyper-parameters are selected via the nested $5$-fold cross-validation and random search. 
Visibly, RoBERTa outperforms all other models in both macro accuracy and macro F1 scores. 
Therefore, in the rest of this paper, we employ RoBERTa for classifying and sampling unlabeled data.

\subsubsection{Iteration setups.}
The test dataset $X_{test}$ used for evaluation contains $114$ labeled Facebook postings.
$X_{test}$ is only used in the convergence evaluation and is kept fixed between iterations.
To keep bounded the human annotation effort, we limit each iteration to $100$ postings.
For each of the four topic, we sample $|X_i|=|X_i^{A}| + |X_i^{T}| + |X_i^{R}| = 10 + 10 + 5= 25$ posts from $U_{i-1}$.
Note that $X_i^{A}$, $X_i^{T}$ and $X_i^{R}$ are the sets of samples selected at iteration $i$ using the three strategies introduces in \cref{subsec:dataset-augmentation}.
Also note that identical postings may be selected multiple times for different topics.

In total, we conduct $7$ iterations of augmentation until we observe convergence in classification performance on $X_{test}$ (see convergence analysis in \Cref{subsec:augmentation-results}).
The first $4$ iterations sampled only Facebook postings as this is the prominent source in our dataset and most used social media in Australia \citep{newman2020reuters}.
After the $5$th iteration, we introduced the other two data sources, Twitter and Youtube.

\subsection{Augmentation Results}
\label{subsec:augmentation-results}

\subsubsection{Augmented dataset statistics.} 
\Cref{tab:dataset} compares the number of postings and opinions between the dataset constructed by the qualitative analysis ($L_0$) and the final labeled dataset after the seventh iteration ($L_7$).
$L_7$ contains $1,381$ postings and $71$ opinions, which is more than double those of $L_0$ ($614$ postings and $65$ opinions). 
We note that \textit{Climate change} is the most prevalent topic in the dataset ($592$ postings in $L_{7}$) while \textit{Austalian bushfire} is the least ($287$ postings).

\subsubsection{Emergence of new opinions.}
During the data augmentation process, the experts continuously evolved the opinion set in addition to labeling new data.
For example, opinions such as ``Covid-19 is a plague sent by God'' were detected and reinforced by the data sampling strategies. 
Similarly, the data sampled uncovered a longer duration of opinions than the range explored by the experts in the qualitative study.
These provided the qualitative researchers with a long-term perspective about how opinions emerge temporally (see more detailed analysis in \cref{sec:opinion-analysis}).
Overall, \cref{tab:dataset} shows that $6$ new opinions have emerged between $L_0$ and $L_7$. 
We refer to \citep{appendix} for a complete list of opinions and their volumes at $L_0$ and $L_7$.

\begin{figure}[!tbp]
	\centering
	\begin{subfigure}{0.49\textwidth}
		\includegraphics[page=2,width=\textwidth]{images/classifiers_at_batches.pdf}
		\caption{}
	\end{subfigure}
	\begin{subfigure}{0.49\textwidth}
		\includegraphics[width=\textwidth,page=1]{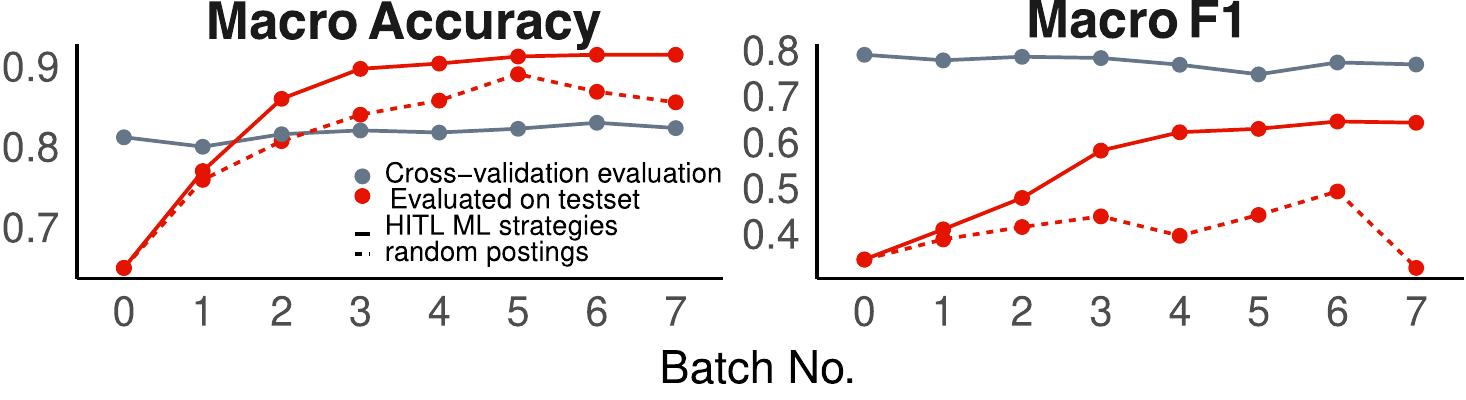}
		\caption{}
		\label{subfig:macro_classifier_convergence}
	\end{subfigure}
	\caption{
        Convergence of topic classifier performances over seven iterations.
		\textbf{(a)} Evaluation by topics on test set and \textbf{(b)} macro-aggregated over all topics on test set and via cross-validation.
		The solid lines show our proposed HITL data augmentation strategy, while the dashed lines show the random selection baseline.
	}
	\label{fig:classifier_convergence}
\end{figure}

\subsubsection{Convergence analysis.}
\Cref{fig:classifier_convergence}a shows the prediction performance on the test set $X_{test}$, for each topic (accuracy on the left panel, and F1 score on the right panel), over iterations $0$ to $7$.
The solid lines in \Cref{fig:classifier_convergence}b show the performance indicators macro-averaged over topics, together with the cross-validation generalization error (see the iterations and convergence discussion in \Cref{subsec:dataset-augmentation}).

All indicators show that prediction performance improves over subsequent iterations, with the topic \textit{2019-20 Australian bushfire season} demonstrating the fastest growth.
Both accuracy and F1 scores on the test data converge fast in the first $3-4$ iterations, while improvements from the subsequent iterations are limited.
This suggests a reduced marginal utility of the later iterations.
Notably, the performance gain is null between iterations 6 and 7, suggesting that the procedure has converged.
Consequently, we stopped the data augmentation process after the seventh iteration.

The cross-validation performance is stable across iterations.
This is expected as the classifiers learn from the same data on which the generalization is estimated --- i.e., the classifiers are representative of the data they were trained on.
However, the difference between the test set performance and cross-validation performance is indicative of the representativity over the entire dataset which improves as more iterations are performed.
The cross-validation accuracy is consistently lower than the test set accuracy because the test data is more imbalanced than labeled data.
The cross-validation F1 is more optimistic than the test set F1.
Finally, the difference between the two stabilizes for the later iterations, further suggesting the convergence.

\subsubsection{Baseline comparison.} 
We compare the sampling strategies defined in \cref{subsec:dataset-augmentation} with a baseline scenario where we code an equal amount of postings that were all randomly sampled. 
This results in a sequence of baseline batches of postings which are manually annotated by the experts using the exact same procedure as before.
Next, we train classifiers with these baseline batches in iterations and compute the prediction performance on the same test set.
\Cref{subfig:macro_classifier_convergence} shows the baseline performance as dashed lines.
Visibly, the macro accuracy and F1 scores show increasing gaps between the proposed method and the baseline labeling scenario.
This indicates the advantage of our chosen data augmentation strategies, particularly the active learning strategy which is known to outperform random sampling~\citep{tran2018combining}.

\section{Case Studies}
\label{sec:case_studies}
In this section, we present a case study of Facebook posts from an Australian public page.
The page shifts between early 2020 (\emph{2019-2020 Australian bushfire season}) and late 2020 (\emph{COVID-19 crises}) from being a moderate-right group for discussion around climate change to a far-right extremist group for conspiracy theories.

\begin{figure*}[!tbp]
	\begin{subfigure}{0.21\textwidth}
		\includegraphics[width=\textwidth]{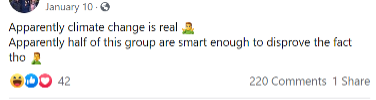}
		\caption{}
		\label{subfig:first-posting}
		\includegraphics[width=0.9\textwidth]{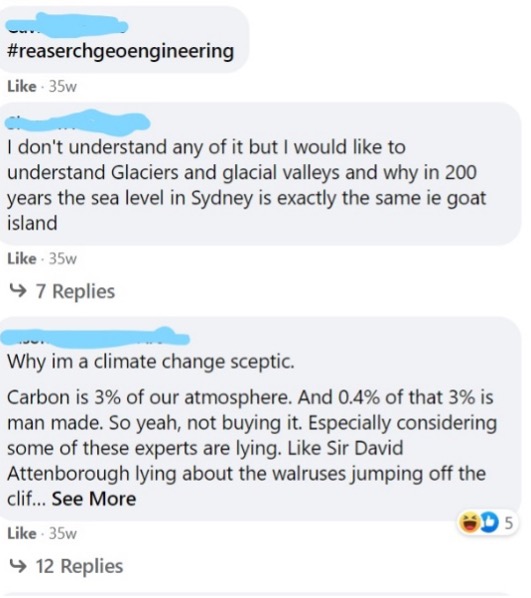}
		\caption{}
		\label{subfig:comment-post-1}
	\end{subfigure}
    \begin{subfigure}{0.28\textwidth}
		\includegraphics[width=\textwidth]{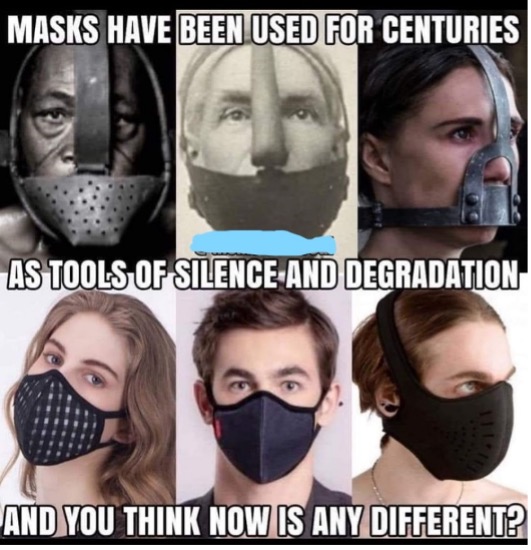}
		\caption{}
		\label{subfig:second-posting}
	\end{subfigure}
    \begin{subfigure}{0.23\textwidth}
		\includegraphics[width=\textwidth]{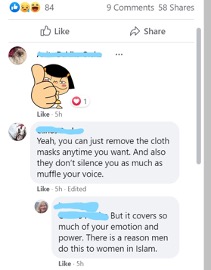}
		\caption{}
		\label{subfig:comment-post-2a}
	\end{subfigure}
    \begin{subfigure}{0.23\textwidth}
		\includegraphics[width=\textwidth]{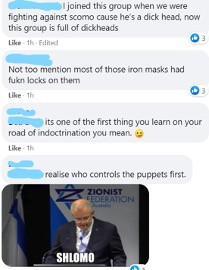}
		\caption{}
		\label{subfig:comment-post-2b}
	\end{subfigure}
	\caption{
		Examples of postings and comment threads from a public Facebook page from two periods of time early 2020 (a) and late 2020 (b)-(e), which show a shift from climate change debates to extremist and far-right messaging.
	}
	\label{fig:facebook}
\end{figure*}

We focus on a sample of 2 postings and commenting threads from one Australian Facebook page we classified as ``far-right'' based on the content on the page. 
We have anonymized the users in \Cref{fig:facebook} to avoid re-identification.
The first posting and comment thread (see \Cref{subfig:first-posting}) was collected on Jan 10, 2020, and responded to the Australian bushfire crisis that began in late 2019 and was still ongoing in January 2020. It contains an ambivalent text-based provocation that references disputes in the community regarding the validity of climate change and climate science. 

The second posting and comment thread (see \Cref{subfig:second-posting}) was collected from the same page in September 2020, months after the bushfire crisis had abated.
At that time, a new crisis was energizing and connecting the far-right groups in our dataset --- i.e., the COVID-19 pandemic and the government interventions to curb the spread of the virus. 
The post is different in style compared to the first.
It is image-based instead of text-based and highly emotive, with a photo collage bringing together images of prison inmates with iron masks on their faces (top row) juxtaposed to people wearing face masks during COVID-19 (bottom row). 
The image references the public health orders issued during Melbourne's second lockdown and suggests that being ordered to wear masks is an infringement of citizen rights and freedoms, similar to dehumanizing restraints used on prisoners.

To analyze reactions to the posts, two researchers used a deductive analytical approach to separately code and to analyze the commenting threads --- see \Cref{subfig:comment-post-1} for comments of the first posting, and \Cref{subfig:comment-post-2a,subfig:comment-post-2b} for comments on the second posting. 
Conversations were also inductively coded for emerging themes. 
During the analysis, we observed qualitative differences in the types of content users posted, interactions between commenters, tone and language of debate, linked media shared in the commenting thread, and the opinions expressed.
The rest of this section further details these differences.
To ensure this was not a random occurrence, we tested the exemplar threads against field notes collected on the group during the entire study.
We also used Facebook's search function within pages to find a sample of posts from the same period and which dealt with similar topics. 
After this analysis, we can confidently say that key changes occurred in the group between the bushfire crisis and COVID-19, that we detail next.

\subsubsection*{Exemplar 1 --- climate change skepticism.}
To explore this transformation in more depth, we analyzed comments scraped on the first posting --- \cref{sub@subfig:comment-post-1} shows a small sample of these comments.
The language used was similar to comments that we observed on numerous far-right nationalist pages at the time of the bushfires.
These comments are usually text-based, employing emojis to denote emotions, and sometimes being mocking or provocative in tone. 
Noteworthy for this commenting thread is the 50/50 split in the number of members posting in favor of action on climate change (on one side) and those who posted anti-Greens and anti-climate change science posts and memes (on the other side).
The two sides aligned strongly with political partisanship --- either with Liberal/National coalition (climate change deniers) or Labor/Green (climate change believers) parties. 
This is rather unusual for pages classified as far-right. 

We observed trolling practices between the climate change deniers and believers, which often descend into \emph{flame wars} --- i.e., online ``firefights that take place between disembodied combatants on electronic bulletin boards''~\citep{bukatman1994flame}.
The result is a boosted engagement on the post but also the frustration and confusion of community members and lurkers who came to the discussions to become informed or debate rationally on key differences between the two positions.
They often even become targeted, victimized, and baited by trolls on both sides of the partisan divide. 
The opinions expressed by deniers in commenting sections range from skepticism regarding climate change science to plain denial.
Deniers also regard a range of targets as embroiled in a climate change conspiracy to deceive the public, such as The Greens and their environmental policy, in some cases the government, the United Nations, and climate change celebrities like David Attenborough and Greta Thunberg. 
These figures are blamed for either exaggerating risks of climate change or creating a climate change hoax to increase the influence of the UN on domestic governments or to increase domestic governments' social control over citizens. 

Both coders noted that flame wars between these opposing personas contained very few links to external media. 
Where links were added, they often seemed disconnected from the rest of the conversation and were from users whose profiles suggested they believed in more radical conspiracy theories.
One such example is ``geo-engineering'' (see \cref{sub@subfig:comment-post-1}).
Its adherents believe that solar geo-engineering programs designed to combat climate change are secretly used by a global elite to depopulate the world through sterilization or to control and weaponize the weather.

Nonetheless, apart from the random comments that hijack the thread, redirecting users to external ``alternative'' news sites and Twitter, and the trolls who seem to delight in victimizing unsuspecting victims, the discussion was pretty healthy.
There are many questions, rational inquiries, and debates between users of different political persuasion and views on climate change.
This, however, changes in the span of only a couple of months.

\subsubsection*{Exemplar 2 --- posting and commenting thread.}
We observe a shift in the comment section of the post collected during the second wave of the COVID pandemic (\Cref{sub@subfig:second-posting}) --- which coincided with government laws mandating the public to wear masks and stay at home in Victoria, Australia.
There emerges much more extreme far-right content that converges with anti-vaccination opinions and content.
We also note a much higher prevalence of conspiracy theories often implicating racialized targets.
This is exemplified in the comments on the second post (\Cref{sub@subfig:comment-post-2a,sub@subfig:comment-post-2b}) where Islamophobia and antisemitism are confidently asserted alongside anti-mask rhetoric.
These comments consider face masks similar to the religious head coverings worn by some Muslim women, which users describe as ``oppressive'' and ``silencing''. 
In this way, anti-maskers cast women as a distinct, sympathetic marginalized demographic.
However, this is enacted alongside the racialization and demonization of Islam as an oppressive religion. 

Given the extreme racialization of anti-mask rhetoric, some commenters contest these positions, arguing that the page is becoming less an anti-Scott Morrison page (Australia's Prime Minister at the time) and changing into a page that harbors ``far-right dickheads''.
This questioning is actively challenged by far-right commenters and conspiracy theorists on the page, who regarded pro-mask users and the Scott Morrison government as ``puppets'' being manipulated by higher forces (see \Cref{sub@subfig:comment-post-2b}). 

This indicates a significant change on the page's membership towards the extreme-right, who employs more extreme forms of racialized imagery, with more extreme opinion being shared.
Conspiracy theorists become more active and vocal, and they consistently challenge the opinions of both center conservative and left-leaning users. 
This is evident in the final two comments in \Cref{subfig:comment-post-2b}, which reflect QAnon style conspiracy theories and language.
Public health orders to wear masks are being connected to a conspiracy that all of these decisions are directed by a secret network of global Jewish elites, who manipulate the pandemic to increase their power and control. 
This rhetoric intersects with the contemporary ``QAnon'' conspiracy theory, which evolved from the ``Pizzagate'' conspiracy theory.
They also heavily draw on well-established antisemitic blood libel conspiracy theories, which foster beliefs that a powerful global elite is controlling the decisions of organizations such as WHO and are responsible for the vaccine rollout and public health orders related to the pandemic.
The QAnon conspiracy is also influenced by Bill Gates' Microchips conspiracy theory, i.e., the theory that the WHO and the Bill Gates Foundation global vaccine programs are used to inject tracking microchips into people.

These conspiracy theories have, since COVID-19, connected formerly separate communities and discourses, uniting existing anti-vaxxer communities, older demographics who are mistrustful of technology, far-right communities suspicious of global and national left-wing agendas, communities protesting against 5G mobile networks (for fear that they will brainwash, control, or harm people), as well as generating its own followers out of those anxious during the 2020 onset of the COVID-19 pandemic.
We detect and describe some of these opinion dynamics in the next section.

\section{Opinion Dynamics and Network Centrality}
\label{sec:opinion-analysis}

This section first examines the relative importance of opinions in online discussions, obtained from a large sample of machine-labeled postings.
This allows the application of the qualitative-defined coding schema (see \cref{subsec:qual-study}) to a significantly larger sample of postings, reducing the unavoidable selection bias of the qualitative study. 
Next, we study the dynamics of opinion co-occurrences. We note that, due to large overlaps in posting times and similarities in topics, the analysis of opinions in this section is conducted on two topic groups: \textit{2019-20 Australian bushfire season}, \textit{climate change}, and \textit{Covid-19}, \textit{vaccination} (also shown in \Cref{tab:keywords}).

\begin{figure}[!tbp]
	\centering
	\includegraphics[width=0.48\textwidth]{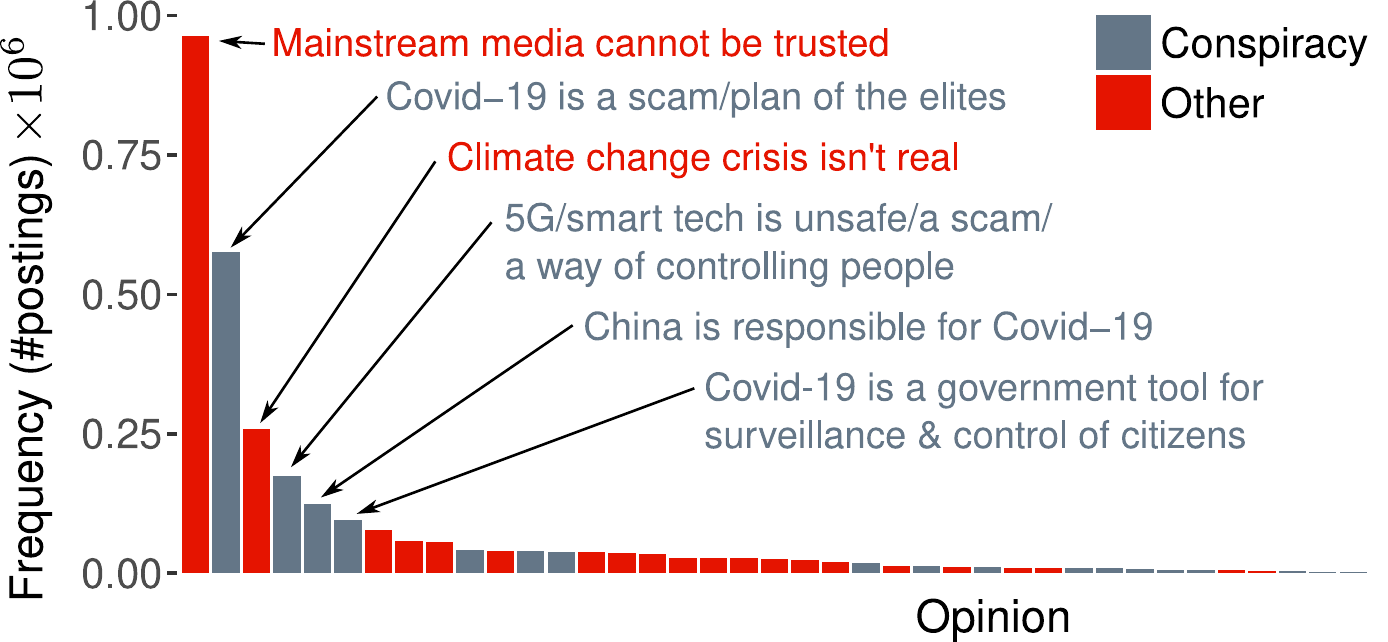}
	\caption{
        The usage frequency of each of opinions in a large sample of machine-labeled data shows a long-tail distribution.
        Four of the top six opinions endorse a conspiracy theory (shown in gray).
	}
	\label{fig:opinion-n}
\end{figure}
\noindent {\bf Experimental setups.}
After completing the last iteration of the dataset augmentation ($L_7$), we train the topic and opinion classifiers (see \Cref{subsec:dataset-augmentation}) on all available training data.
We apply these classifiers to all available unlabelled samples ---
$22,965,816$ postings in total.
The vast majority of these ($21,266,038$) are off-topic, i.e., with no opinion associated. 
This is expected given the broad keyword sampling of our unlabeled dataset.
The remainder of $1,699,778$ postings are labeled with at least one opinion, and $313,720$ postings were associated with more than one opinion.
This creates $2,089,336$ posting-opinion relations, which we use in the rest of this section to analyze the dynamics of opinions. 
We manually identify the opinion labels that relate to \textit{conspiracy theories}
and we discuss them in the experimental results.
We show in the appendix~\citep{appendix} the complete list of opinions and those relating to conspiracy theories.

\subsection{Opinion Frequency Distribution} %
\label{subsec:frequency-analysis}
We show in \Cref{fig:opinion-n} the frequency distribution of opinions in the machine-labeled data.
Unsurprisingly (in hindsight), the size distribution for opinions is long-tailed, commonly emerging in online measurements.
This translates into a relatively small number of opinions monopolizing the online debate.
Perhaps more surprisingly, most of the prevalent opinions are linked to conspiracy theories; 
four among the top six most popular opinions are conspiracy theories, including 
``Covid-19 is a scam/plan of the elites'' (2nd most frequent opinion),
``5G/smart tech is unsafe/a scam/a way of controlling people'' (4th),
``China is responsible for Covid-19'' (5th), and 
``Covid-19 is a government tool to increase the powers of the state and surveillance/control of citizens'' (6th). 
This showcases the advantages of our mixed-method approach: our qualitative case studies (see \cref{sec:case_studies}) identify the importance of conspiracy theories in the online debate; still, they could not assess the scope of their importance relative to all the other opinions.
We further show in the appendix~\citep{appendix} the daily relative frequency of top opinions.

\subsection{Centrality Dynamics in Opinion Networks}

\begin{figure*}[!tbp]
	\begin{subfigure}{\textwidth}
		\includegraphics[width=\textwidth]{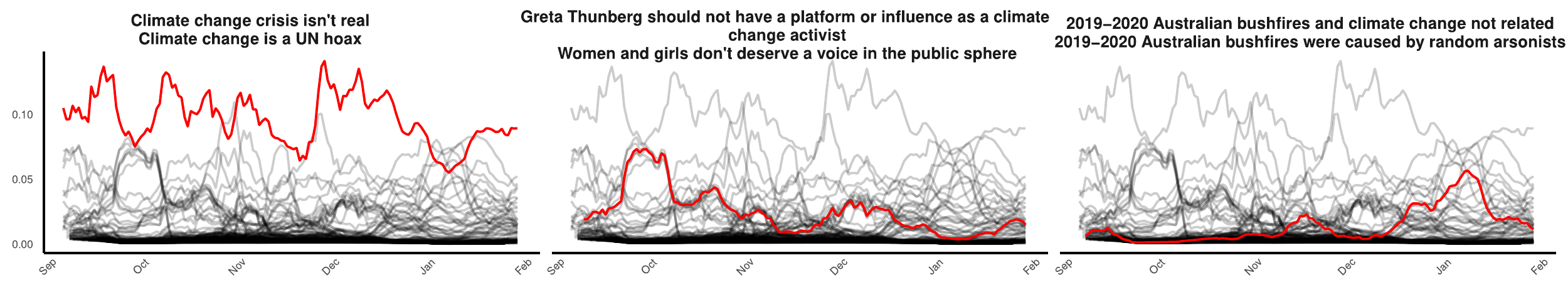}
	\end{subfigure}
	\caption{
		Daily proportions of all edge weights (gray lines) representing co-occurred opinions pairs.
		The red lines show three selected dynamics: 
		continuous strong association (left panel), declining weight (center panel) and increasing weight (right panel).
		At any given time point, the values on all lines sum to one.
	}
	\label{subfig:mapping_1}
\end{figure*}

\begin{figure}[!tbp]
	\centering
	\begin{subfigure}{0.48\textwidth}
		\includegraphics[width=\textwidth]{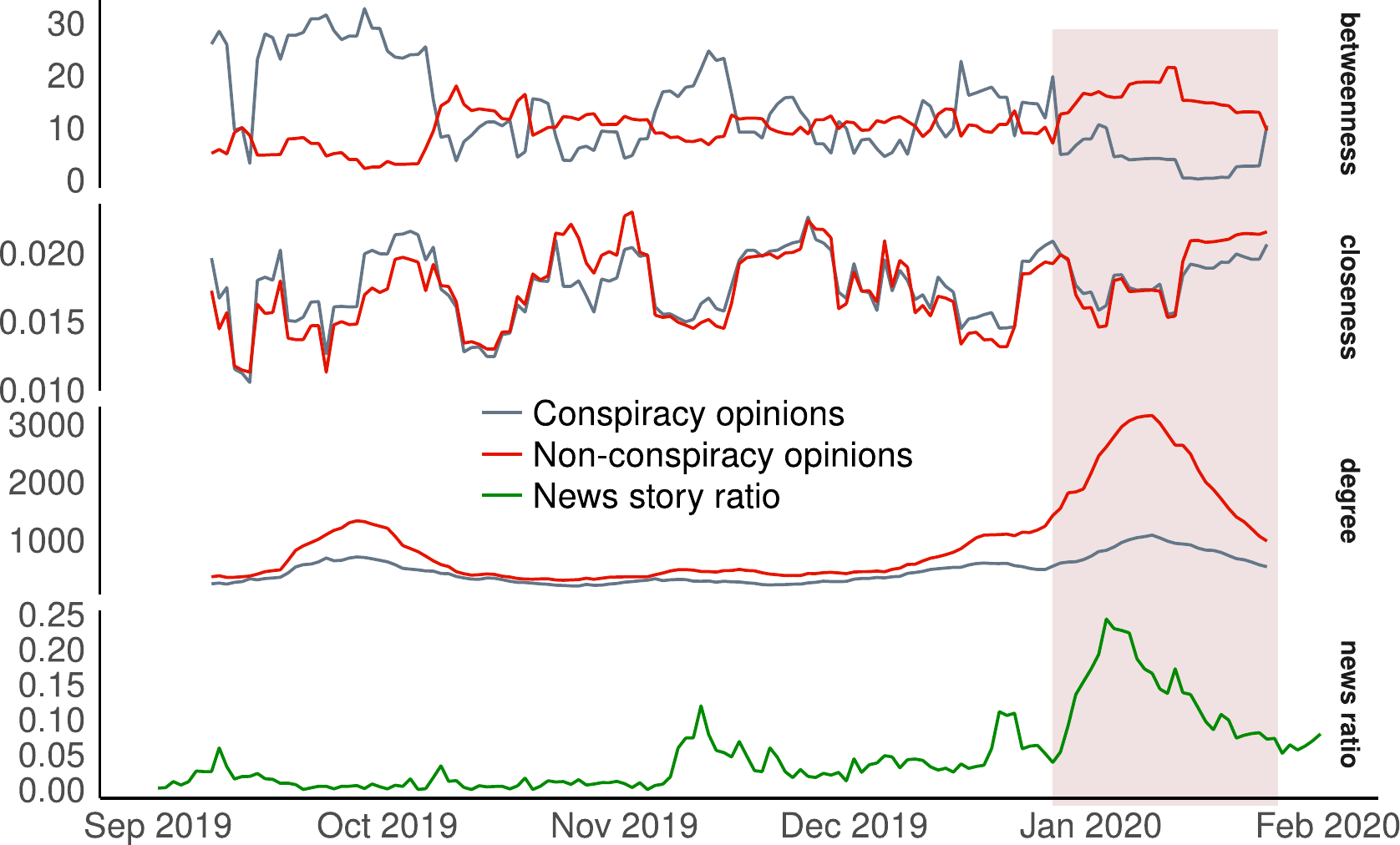}
	\end{subfigure}
	\caption{
		Dynamics of mean centrality measures in the opinion co-occurrence network for conspiracy (red lines) and non-conspiracy opinions (gray lines).
		The green line shows the news coverage ratios from Media Cloud \citep{roberts2021media}.
		The highlighted area shows a spike in the news coverage, which coincides with a decrease in the centrality of conspiracy opinions.
	}
	\label{subfig:mapping_2}
\end{figure}
\begin{figure}[!tbp]
		\begin{subfigure}{0.49\textwidth}
		\includegraphics[width=\textwidth]{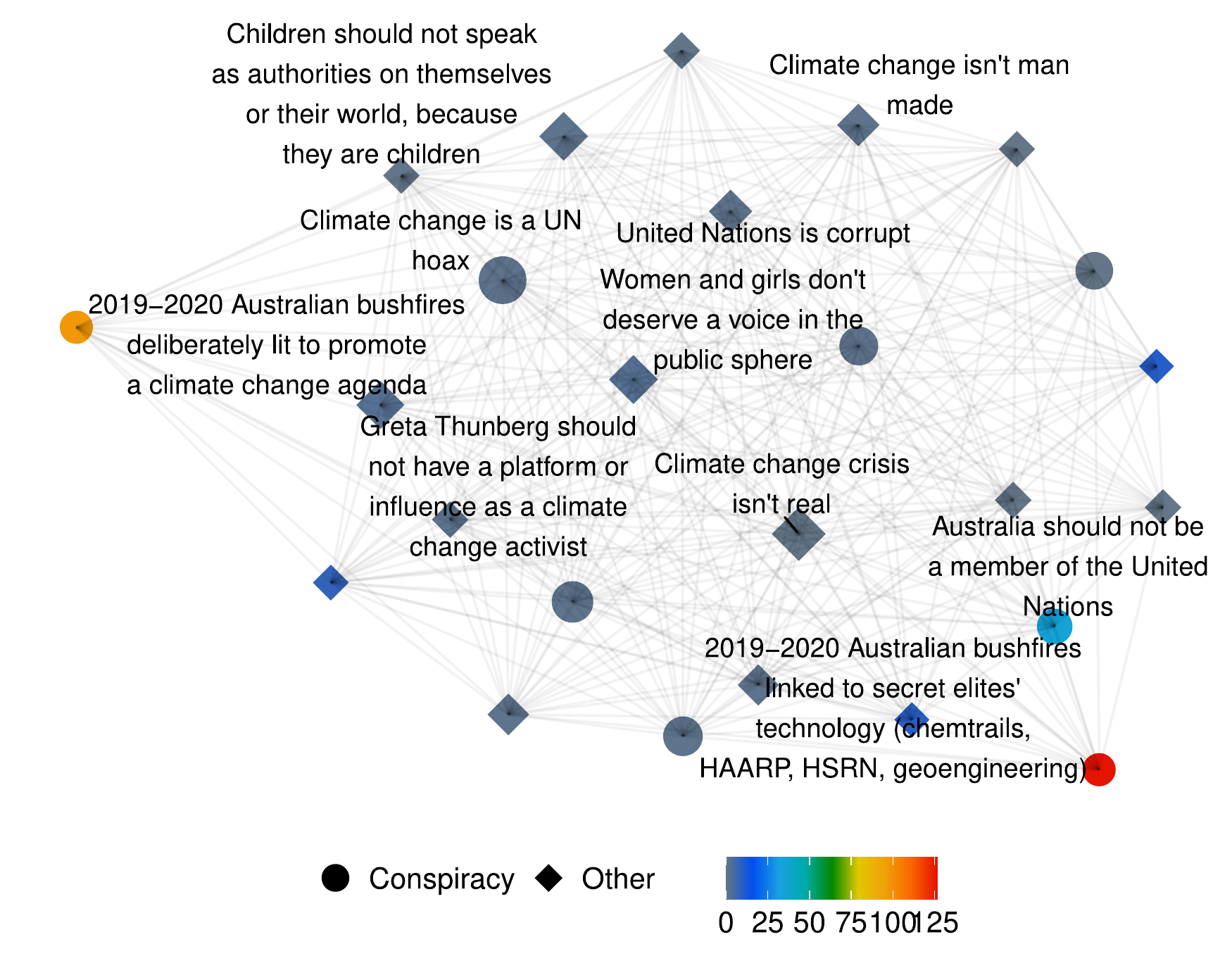}
	\end{subfigure}
	\caption{A visualization of the co-occurrence network in late September 2020 --- node sizes and colors indicate the degrees and betweenness values, respectivelly.}
	\label{subfig:mapping_3}
\end{figure}

\textbf{Build the opinion co-occurrence network.}
It is common that postings express multiple opinions. 
Such co-occurring opinions help identify central opinions, which usually spawn new emerging opinions. 
Here we investigate this process by building the opinion co-occurrence network in the online conversation of the topic \textit{2019-20 Australian bushfire season}.
In the network, the nodes represent the 27 opinions captured during the bushfire conversation.
An edge between two nodes exists when both opinions are present together in at least one posting. 
The node degree of a given opinion node represents the number of opinions that co-occurred with it. 
The edges are weighted by the number of postings in which their connected node opinions co-occur.

\textbf{Dynamics of topic co-occurrence intensity.}
We first investigate the evolution of opinion co-occurrences. 
In \Cref{subfig:mapping_1}, we plot the daily proportions of weights of each edge among all edges, from September 2019 to January 2020. 
We showcase three selected edges (i.e., opinion pairs) that are representative of three types of temporal dynamics:
\begin{itemize}
    \item A continuous and relatively strong association between prevalent opinions --- ``Climate change crisis isn't real'' and ``Climate change is a UN hoax'', the latter notably being a conspiracy theory. 
    \item Associations with declining relative frequencies --- ``Greta Thunberg should not have a platform or influence as a climate...'' and ``Women and girls don't deserve a voice in the public sphere''.
    \item Rising associations --- ``bushfires and climate change not related'' and ``bushfires were caused by random arsonists''.
\end{itemize}
\rev{These three types of co-occurrence dynamics can inform how potentially harmful opinions are selectively co-used with other opinions, and can serve as early warnings for their adoption (and possibly normalization) by participants.
However, to gain a structural understanding of the role of harmful opinions in the broader debate, we next study the structure and dynamics of the opinion co-occurrence network.}

\textbf{Centrality of conspiracy opinions and news ratio.}
Here, we study the importance of conspiracy opinions over time measured using their centrality in the dynamic network of opinions.
The network is constructed for each day, and an edge exists if the pair of opinions co-occurs at least once.
We measure nodes' centrality using three measures: betweenness, closeness, and node degrees.
\Cref{subfig:mapping_2} shows the average centrality for each measure for the $8$ conspiracy and $19$ non-conspiracy opinions.
We also depict the attention dedicated by the Australian news media to the bushfires during the same period. 
We estimate the latter using the news coverage ratio --- the percentage of articles dedicated to the topic over all captured articles in a day ---
crawled using the Media Cloud \citep{roberts2021media}. 

We observe that the conspiracy opinions have higher mean betweenness than the non-conspiracy opinions in September 2019 and again in November 2019.
It is only in January 2020 that their mean centrality decreases consistently, which, interestingly, corresponds to a significant uptick in the attention given by the media.
\revK{This might suggest that the diffusion of more authoritative content by the news media, together with the participation of their readership, crowded out conspiracy opinions and marginalized their impact.}

\textbf{A launching pad for fringe opinions.}
The episodically high centrality of conspiracy opinions suggests they are selectively used in conjunction with many other opinions.
\revK{We posit that contested conspiracy opinions are leveraged together with more accepted and mainstream opinions to rationalize and popularize them.
Furthermore, they are used with existing conspiracy opinions to amplify the influence.}
We test this hypothesis by mapping, in \Cref{subfig:mapping_3}, the opinion co-occurrence network from posts published over 14 days in late September 2019, i.e., the period when the betweenness for conspiracy opinions is at its peak.
At the network's center lie opinions with both high betweenness and high degree, such as ``United Nations is corrupt'' or ``Climate change isn't real''.
These are long-lasting, general-purpose opinions that we frequently find throughout our dataset.
These are also the backbone on which the conspiracy theories build to increase their presence in the narrative.
We find the closely related and very central ``Climate change is a UN hoax'', but also more fringe opinions towards the periphery of the network --- such as 
``Bushfires linked to secret elites' secret technology (chemtrails, HAARP, HSRN, geoengineering)'', 
``bushfires deliberately lit to promote a climate change agenda'' and 
``Australia should not be a member of the United Nations''.

\section{Related Work}
\textbf{Problematic speech datasets.}
Several datasets \citep{wang2017liar,shu2018fakenewsnet,hasan2020truth} on problematic speeches have been made available recently. Among these, \citet{wang2017liar,shu2018fakenewsnet} crawled and used labels from existing fact-checking sites (e.g., POLITIFACT\footnote{\url{https://www.politifact.com}}), whereas \citet{hasan2020truth} employed an active learning component in their pipeline with a goal of maximizing the accuracy of fake news detection.

\textbf{Human-in-the-loop.}
HITL machine learning algorithms have been widely applied for building datasets in various applications, including sentiment analysis \citep{mozafari2014scaling}, computer vision \citep{vijayanarasimhan2011cost} and medical image classification \citep{hoi2006batch}. 
\citet{wang2021putting} provide a comprehensive review of applying HITL methods to natural language processing tasks, in which they stress the importance of designing both quantitative and qualitative methods to evaluate human feedback for complex feedback types. 
In particular, \citet{chen2018using} propose to identify ambiguity in qualitative coding via active learning, which is the most relevant work to ours. 
In this paper, we extend the method by introducing two other strategies to balance exploration and exploitation.

Overall, our study differs from prior works by highlighting the benefits of deploying HITL algorithms to accelerate qualitative studies on online problematic speeches. The augmented data in this paper exposes us to a richer context of problematic discussions where we can identify trajectories of opinion evolutions (in \Cref{sec:opinion-analysis}).

\section{Conclusion}
This work proposes a solution that fills the gap between qualitative and quantitative analysis of problematic online speech.
We construct an ontology (using Wikibase) which is initially populated through a qualitative study.
The latter emerges from both the vocabulary of annotations (the opinions expressed in topics) and collected labeled data from three online social network platforms (Facebook, Twitter, and Youtube).
Next, we collect a large dataset of social media data using keyword search.
Finally, we augment the labeled dataset using a human-in-the-loop machine learning algorithm. 
We present two in-detail case studies with observations of problematic online speech which evolved on an Australian far-right Facebook group. 
Using our machine-labeled dataset, we analyze how problematic opinions emerge over time and how they co-occur.

\subsubsection{Limitations.}
The present study has several limitations, which we group into data and methodological limitations.

The data limitations are mainly related to the human labeling bias, considered platforms, and posting accessibility.
The initial qualitative study, conducted by the team members, may suffer from human labeling bias.
This is a known limitation of qualitative methods, which we partially alleviate using our data augmentation procedure.
Second, this study concentrates on three platforms (Facebook, Twitter, and Youtube), and Facebook makes most of our data sample.
However, all three are mainstream platforms; problematic speech also occurs outside these platforms, and future work would need to account for platforms like 8chan or gab.
Last, our study only leverages public postings --- we do not access the private conversations for technical and ethical reasons. 

We mention four methodological limitations.
First, the quality of the classifier is inferior to any human coder. 
Yet, this is a marginal problem when the goal is not to correctly label each posting but instead to capture patterns across a large number of postings. 
Second, the definition of the set of Internet sources where the data collection occurs remains critical in determining how representative the sample still is; a larger set of Internet places might not address the selection bias (if they are all selected the same way).
Third, the active learning and top confidence sampling strategies that exploit the labeled dataset may further reinforce the initial human sampling bias. 
We mitigate this shortcoming via random sampling strategy.
Last, by design, the classifiers we have deployed cannot identify opinions that were not identified during the qualitative study.
Future research could apply dynamic predictive models designed to capture the label distribution shift and construct an active set of labels.

\section*{Acknowledgements}
This research was partially funded by the University of Technology of Sydney through a cross-faculty grant, Facebook Research under the Content Policy Research Initiative, and the Commonwealth of Australia (represented by the Defence Science and Technology Group) through a Defence Science Partnerships Agreement.

{\fontsize{9.8pt}{10pt}\selectfont
\bibliography{main}}

\clearpage
\appendix
\section{Opinion Analysis}\label{sec:appendix}

\subsection{Complete Opinion List}
We provide a complete list of opinions obtained from the labeling \revK{iterations} in \Cref{tab:opinion_list} along with the frequencies of individual opinions in $L_0$ and $L_7$.

\subsection{Top Opinion Daily Relative Frequency}
\begin{figure*}[!tbp]
	\centering
    \begin{subfigure}{0.99\textwidth}
		\includegraphics[width=\textwidth,page=1]{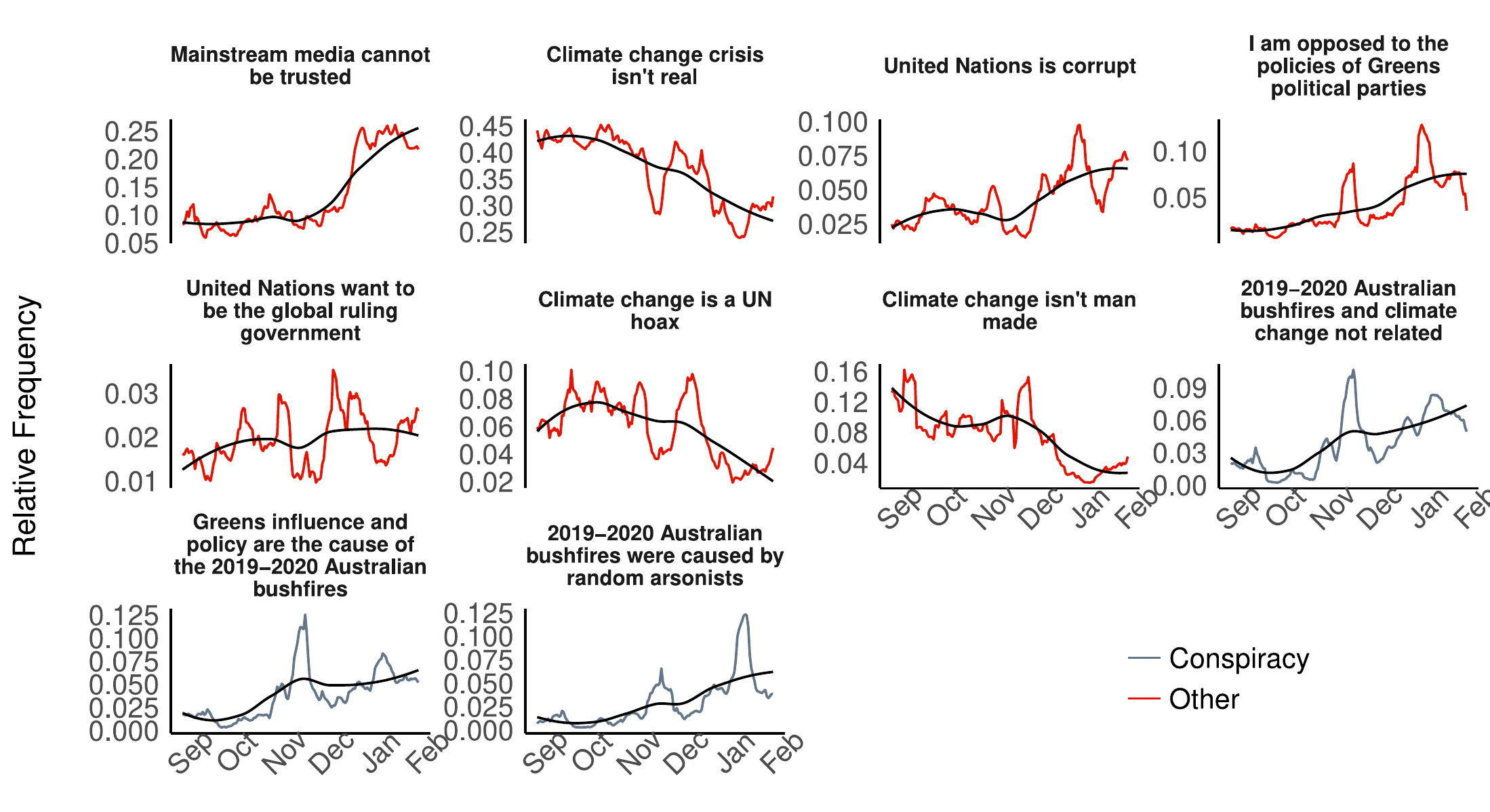}
		\caption{}
		\label{fig:opinion_relative_frequency_1}
	\end{subfigure}
	\begin{subfigure}{0.99\textwidth}
		\includegraphics[width=\textwidth,page=2]{images/opinion_relative_frequency.pdf}
		\caption{}
		\label{fig:opinion_relative_frequency_2}
	\end{subfigure}
	\caption{
		Daily relative frequency of top 10 opinions of (a) \textit{2019-20 Australian bushfire season} and \textit{Climate change}; (b) \textit{Covid-19} and \textit{vaccination}. The black lines depict smoothed trends.
	}
	\label{fig:opinion_relative_frequency}
\end{figure*}

\revK{We further show the daily relative frequencies of top $10$ opinions of \textit{2019-20 Australian bushfire season}, \textit{climate change} in \Cref{fig:opinion_relative_frequency_1} and \textit{Covid-19}, \textit{vaccination} in \Cref{fig:opinion_relative_frequency_2}.
For both groups of plots, notably, general upward trends over times are observed for conspiracy opinions indicating an increase of attention from users, while more downward trends are prevalent for other opinions.}

\clearpage
\begin{table*}
\caption{\label{tab:opinion_list}List of opinions and their frequencies. The conspiracy opinions are in bold text.}

\centering
\scriptsize
\begin{tabular}{p{8cm}p{4cm}rr}
\toprule
opinion & topic & $L_0$ & $L_7$\\
\midrule
Climate change crisis isn't real & climate change & 113 & 252\\
\textbf{Covid-19 is a scam/plan of the elites} & Covid-19 & 111 & 152\\
United Nations is corrupt & climate change & 75 & 110\\
\textbf{Climate change is a UN hoax} & climate change & 72 & 104\\
\textbf{Australia should not be a member of the United Nations} & climate change & 66 & 49\\
\addlinespace
Mainstream media cannot be trusted & climate change, vaccination, 2019-20 Australian bushfire season, Covid-19 & 50 & 100\\
\textbf{United Nations want to be the global ruling government} & climate change & 42 & 66\\
\textbf{Covid-19 is a government tool to increase the powers of the state and surveillance/control of citizens} & vaccination, Covid-19 & 34 & 43\\
Vaccination sacrifices bodily autonomy & vaccination, Covid-19 & 33 & 30\\
2019-2020 Australian bushfires were caused by random arsonists & climate change, 2019-20 Australian bushfire season & 31 & 42\\
\addlinespace
\textbf{China is responsible for Covid-19} & Covid-19 & 28 & 49\\
I am opposed to the policies of Greens political parties & climate change, 2019-20 Australian bushfire season & 26 & 35\\
Vaccines contain dangerous ingredients or are manufactured unsafely & vaccination & 26 & 28\\
\textbf{Vaccines are a government/""elites"" tool to track and control people's behaviour} & vaccination, Covid-19 & 24 & 29\\
Greens influence and policy are the cause of the 2019-2020 Australian bushfires & climate change, 2019-20 Australian bushfire season & 22 & 37\\
\addlinespace
\textbf{Covid-19 is not a real/serious illness} & Covid-19 & 22 & 27\\
\textbf{Vaccine manufacturers cannot be trusted} & vaccination & 22 & 33\\
Changes in the earth's climate are a natural, normal phenomenon & climate change, 2019-20 Australian bushfire season & 21 & 38\\
Climate change isn't man made & climate change, 2019-20 Australian bushfire season & 20 & 45\\
Wellness and alternative health is superior to medicine and science & vaccination & 17 & 17\\
\addlinespace
2019-2020 Australian bushfires and climate change not related & climate change, 2019-20 Australian bushfire season & 16 & 23\\
\textbf{Covid-19 is part of Bill Gates' plan for \#DigitalIDs (\#ID2020)} & vaccination, Covid-19 & 14 & 21\\
Children should not speak as authorities on themselves or their world, because they are children & climate change & 13 & 31\\
Countries that contribute more carbon emissions should make environmental/policy changes, not Australia & climate change & 13 & 14\\
\textbf{2019-2020 Australian bushfires linked to secret elites' technology (chemtrails, HAARP, HSRN, geoengineering)} & climate change, 2019-20 Australian bushfire season & 13 & 11\\
\addlinespace
Vaccines are not safe for children & vaccination & 13 & 18\\
\textbf{Experts manipulate data for private or corporate agendas} & climate change & 13 & 30\\
\textbf{Covid-19 is the Chinese government's bioweapon} & Covid-19 & 13 & 21\\
\textbf{2019-2020 Australian bushfires deliberately lit to promote a climate change agenda} & climate change, 2019-20 Australian bushfire season & 13 & 17\\
Bush fires are a normal summer occurrence for Australia & climate change, 2019-20 Australian bushfire season & 12 & 14\\
\addlinespace
Vaccines are ineffective or unreliable & vaccination & 12 & 14\\
\textbf{Climate change is a scam to generate profit for the wealthy and powerful} & climate change & 11 & 34\\
Economic impacts are the worst outcome and should be avoided at all costs & climate change, Covid-19 & 11 & 21\\
China's government/institutions have too much influence in/on Australia & Covid-19 & 11 & 38\\
Greta Thunberg should not have a platform or influence as a climate change activist & climate change & 9 & 25\\
\addlinespace
\textbf{Covid-19 was developed for profit} & vaccination, Covid-19 & 9 & 10\\
\bottomrule
\end{tabular}
\end{table*}

\begin{table*}
\begin{tabular}{r}
\toprule

\bottomrule
\end{tabular}
\centering
\scriptsize
\begin{tabular}{p{8cm}p{4cm}rr}
\toprule
opinion & topic & $L_0$ & $L_7$\\
\midrule
Vaccines cause Autism & vaccination & 9 & 8\\
Covid-19 is comparable to HIV/AIDS & Covid-19 & 9 & 6\\
The World Health Organization is corrupt & vaccination, Covid-19 & 9 & 19\\
Abortion is wrong & vaccination & 9 & 11\\
The democrats caused this, but Trump is fixing it & Covid-19 & 8 & 16\\
\addlinespace
Modern medicine is a corrupting force & vaccination & 8 & 9\\
Universities are corrupt, cashed up institutions suppressing free speech and promoting left-wing ideology & Covid-19 & 8 & 22\\
Autism is bad & climate change, vaccination & 7 & 8\\
\textbf{Men are being chemically emasculated by the government/science/elites} & vaccination, Covid-19 & 7 & 6\\
I have the right to my freedom of speech & vaccination, Covid-19 & 7 & 13\\
\addlinespace
\textbf{Christianity and Christians are being targeted or persecuted by government policy} & Covid-19 & 7 & 12\\
Australia should not send foreign aid to other countries & climate change, 2019-20 Australian bushfire season & 6 & 8\\
\textbf{5G/smart tech is unsafe/a scam/a way of controlling people} & vaccination, Covid-19 & 6 & 25\\
Silent majority supports my views/actions & vaccination & 5 & 6\\
\textbf{WHO enabled China's government in spreading or covering up covid-19} & Covid-19 & 5 & 13\\
\addlinespace
I'm sceptical of mainstream media information & Covid-19 & 5 & 5\\
Left-wing policies/politics/perspectives reduce quality of life (and therefore aren't worthwhile) & climate change, Covid-19 & 5 & 12\\
Indigenous knowledge and practices could have prevented the 2019-2020 Australian bushfires & climate change, 2019-20 Australian bushfire season & 5 & 5\\
\textbf{Daniel Andrews is in league with China/the CCP} & Covid-19 & 4 & 9\\
Overpopulation is the real threat, regardless of whether climate change is real & climate change & 4 & 5\\
\addlinespace
Vitamin C can prevent or cure covid-19 & vaccination, Covid-19 & 3 & 4\\
Left-wing policies/politics/perspectives exclude or harm “poor people” & climate change, Covid-19 & 3 & 5\\
Wombats protected other animals during the 2019-2020 Australian bushfires & climate change, 2019-20 Australian bushfire season & 2 & 2\\
HIV/AIDs is consensual (unlike covid-19) & Covid-19 & 2 & 2\\
\textbf{Elites control the climate and weather with secret technologies (geo-engineering, etc)} & climate change, 2019-20 Australian bushfire season & 2 & 12\\
\addlinespace
Climate change is real but the impact is being overstated & climate change, 2019-20 Australian bushfire season & 2 & 9\\
Minority \#1 is bad, because they discriminate against minority \#2, which I support & Covid-19 & 2 & 3\\
Past climate change predictions have failed, therefore we can't trust current climate change predictions & climate change & 2 & 5\\
\textbf{2019-2020 Australian bushfires were an Islamic State terrorist attack} & 2019-20 Australian bushfire season & 2 & 2\\
Women and girls don't deserve a voice in the public sphere & climate change, Covid-19 & 0 & 46\\
\addlinespace
\textbf{Covid-19 is a plague sent by (the Christian) God} & vaccination, Covid-19 & 0 & 7\\
\textbf{Jewish people are controlling or seeking to control the world} & climate change, vaccination, Covid-19 & 0 & 9\\
\textbf{Covid-19 and its vaccine are tools to depopulate the earth} & vaccination, Covid-19 & 0 & 15\\
\textbf{Vaccines for Covid-19 will genetically alter people's DNA} & vaccination, Covid-19 & 0 & 2\\
Vaccines for Covid-19 contain aborted fetus tissue & vaccination, Covid-19 & 0 & 1\\
\bottomrule
\end{tabular}
\end{table*}

\end{document}